\documentclass{aip-cp}

\usepackage[numbers]{natbib}
\usepackage{graphicx}
\usepackage{enumerate}
\usepackage[english]{babel}
\usepackage[T1]{fontenc}
\usepackage[utf8]{inputenc}
\usepackage{epstopdf}

\newcommand{\hess}{H.E.S.S.}
\newcommand{\fermilat}{\textit{Fermi}-LAT}
\newcommand{\fermi}{\textit{Fermi}}
\newcommand{\magic}{MAGIC}
\newcommand{\veritas}{VERITAS}
\newcommand{\atom}{ATOM}
\newcommand{\threec}{3C~279}
\newcommand{\pks}{PKS~0736+017}

\begin{document}

\title{Target of Opportunity Observations of Blazars with \hess}

\author[aff1]{M. Cerruti\corref{cor1}}
\author[aff2]{M.~Böttcher}
\author[aff3]{N.~Chakraborty}
\author[aff2,aff4]{I.~D.~Davids}
\author[aff5]{M.~F\"u{\ss}ling}
\author[aff6]{F.~Jankowsky}
\author[aff1]{J.~P.~Lenain}
\author[aff7]{M.~Meyer}
\author[aff8]{H.~Prokoph}
\author[aff6]{S.~Wagner}
\author[aff9]{D.~Zaborov}
\author[aff6]{M.~Zacharias} 
\author{the H.E.S.S. Collaboration}

\affil[aff1]{Sorbonne Universités, UPMC, Université Paris Diderot, Sorbonne Paris Cité, CNRS, LPNHE,\\ 4 place Jussieu, F-75252, Paris Cedex 5, France}
\affil[aff2]{Centre for Space Research, North-West University, Potchefstroom 2520, South Africa}
\affil[aff3]{Alexander von Humboldt Fellow at Max-Planck-Institut für Kernphysik, P.O. Box 103980, D
69029 Heidelberg, Germany}
\affil[aff4]{University of Namibia, Department of Physics, Private Bag 13301, Windhoek, Namibia}
\affil[aff5]{DESY, Platanenallee 6, 15738 Zeuthen, Germany}
\affil[aff6]{Landessternwarte, Universit\"at Heidelberg, K\"onigstuhl, D 69117 Heidelberg, Germany}
\affil[aff7]{Oskar Klein Centre, Department of Physics, Stockholm University, Albanova University Center, SE-10691 Stockholm, Sweden}
\affil[aff8]{Department of Physics and Electrical Engineering, Linnaeus University,  351 95 V\"axj\"o, Sweden}
\affil[aff9]{Laboratoire Leprince-Ringuet, Ecole Polytechnique, CNRS/IN2P3, F-91128 Palaiseau, France}

\corresp[cor1]{matteo.cerruti@lpnhe.in2p3.fr}

\maketitle

\begin{abstract}
The very high energy (VHE, E $>$ 100 GeV) sky is dominated by blazars, radio-loud active galactic nuclei whose relativistic jet is closely aligned with the line of sight. Blazars are characterized by rapid variability at all wavelengths and thus an important part of the \hess\ blazar program is devoted to target of opportunity (ToO) observations. \hess\ triggers blazar ToOs on the basis of publicly available blazar observations at longer wavelengths (optical, X-rays, and
$\gamma$-rays), from private optical observations with the ATOM telescope, and from private communications by $\gamma$-ray partners in
the context of MoUs. In 2015, about 70 hours of H.E.S.S. data were taken in the form of blazar ToOs, which represents ~15$\%$ of all extragalactic observations. In this contribution, we present the \hess\ blazar ToO status, and we focus on two major results from the 2015 season: the detection of VHE emission from \threec\ during the June 2015 flare, and the discovery of \pks\ as a new VHE quasar.\\ 
\end{abstract}

\section{INTRODUCTION}
The extragalactic sky at very-high-energies (VHE; E $>$ 100 GeV), as observed from the ground with Cherenkov telescopes such as the High-Energy Stereoscopic System  (\hess), is dominated by blazars, a class of active galactic nuclei (AGN). Observationally, blazars are AGN characterized by non-thermal emission from radio to $\gamma$-rays, rapid variability, and polarized emission in optical. They are considered as radio-loud AGN with the relativistic jet pointing in the direction of the Earth, resulting in the
boosted photon emission dominating over all other AGN components (i.e. the accretion disk, or the broad-line region (BLR), or the X-ray corona). The class of blazars is divided into the two subclasses of BL Lacertae objects and Flat-Spectrum Radio-Quasars (FSRQs) according to the presence (in the latter) or absence (in the former) of emission lines from the BLR in the optical spectrum, and thus their relative strength with respect to the non-thermal emission produced in the relativistic jet.\\

The spectral energy distribution (SED) of blazars is comprised of two distinct broad bumps: a low-energy one, peaking in infrared-to-X-rays, and usually ascribed to synchrotron emission by electrons/positrons in the jet; and a high-energy one, peaking in MeV-to-TeV, and ascribed, in leptonic models, to inverse Compton scattering. While FSRQs show, in general, a low frequency synchrotron peak in the infrared, BL Lacertae objects show a variety of peak frequencies, and are further classified in low/intermediate/high-frequency-peaked objects (LBL/IBL/HBL). Thus, observations in a given energy band select preferentially blazars of a given peak frequency, and the VHE extragalactic sky is indeed highly dominated by HBLs. On the other hand, up to now only five VHE FSRQs are known, in a catalog of 62 VHE blazars\footnote{see http://tevcat.uchicago.edu/ for an updated list of TeV sources.}: \threec\ \citep[detected with \magic,][]{MAGIC3C279}, PKS~1222+216 \citep[detected with \magic\ and \veritas,][]{1222magic, 1222veritas}, PKS~1510-089 \citep[detected with \hess\ and \magic,][]{1510hess, 1510magic}, PKS~1441+25 \citep[detected with \magic\ and \veritas,][]{1441veritas, 1441magic} and the currently farthest VHE emitter S3~0218+35 \citep[detected with \magic,][]{0218magic}.\\

Multi-wavelength observations are needed in order to study the blazar physics and build up a snapshot of a blazar SED. The rapid variability (down to the minute timescale) requires that the multi-wavelength campaigns should be as simultaneous as possible. When a blazar enters a flaring state, alerts are thus distributed in order to simultaneously follow the evolution of the flare in different energy bands.\\

Target of Opportunity (ToOs) observations are an important part of the \hess\ blazar observing program. During the 2015 season, \hess\ spent about 70 hours of observations as ToOs on flaring blazars. This represents about 15\% of the extragalactic observations of \hess\ (including all extragalactic targets, not only blazars). Alerts are usually based on:\\
\begin{itemize}
\item Public alerts from MWL partners, distributed via Astronomer's Telegrams, or the Gamma-ray Coordinates Network (GCN)
\item An automatic pipeline for the daily analysis of \fermilat\ data on all the relevant Fermi sources visible from the \hess\ location; data are quickly analyzed using aperture photometry, and a full likelihood analysis is run in case an enhanced flux level is measured from a source
\item Private optical data from the \atom\ telescope, located on the \hess\ site
\item Private alerts from MWL partners in the context of Memoranda of Understanding (mainly colleagues from other $\gamma$-ray experiences such as MAGIC, VERITAS, FACT, HAWC and Fermi-LAT) \\ 
\end{itemize}

We present here two important results from \hess\ ToO observations during the 2015 observing season: the detection of VHE emission from \threec\ (this quasar was never detected at VHE again after the original detection by \magic\ in 2006 and 2007), and the discovery of a new VHE FSRQ, \pks. A significant portion of ToO observing time was spent on the FSRQ PKS 1510-089 and data on this source are presented in a separate contribution \cite{zacharias1510}. It is important to highlight that all the results presented focus on FSRQs, taking advantage of the lower energy threshold provided by the \hess\ phase II. \\

\section{The \hess\ telescopes and the data analysis}

\hess\ is an array of Cherenkov telescopes located in Khomas Highland, Namibia, at 1800m above the sea level. Operating since 2004, the original configuration of four 12-m diameter telescopes has been upgraded in 2012 (\hess\ phase II) with the addition of a fifth 28-m diameter telescope in the center of the array (CT\ 5).\\

The telescopes observe the rapid and faint Cherenkov flashes produced through the interaction of VHE photons with the Earth's atmosphere. A major limiting factor of this detection technique is represented by the showers produced in the interaction of hadronic cosmic rays with the Earth's atmosphere. Photons and hadrons can be discriminated using the width, length
and other characteristics of the shower images. The results presented in this work have been obtained using the \textit{model} analysis \citep{model}, and crosschecked with an independent analysis chain, \textit{ImPACT} \citep{impact}.\\

During \hess\ phase II, data can be analysed in the monoscopic or the stereoscopic configuration: in the monoscopic one, only the information from the large-size telescope is used, while the stereoscopic one uses the information from the full array. The monoscopic data analysis provides a lower threshold \citep{zaborovicrc}, which is more useful for the study of FSRQ, which are known to be characterized by a soft VHE spectrum. In this contribution we thus present the results from monoscopic analysis only.\\

\section{\threec\ : the June 2015 flare }

\begin{figure}
\includegraphics[width=300pt]{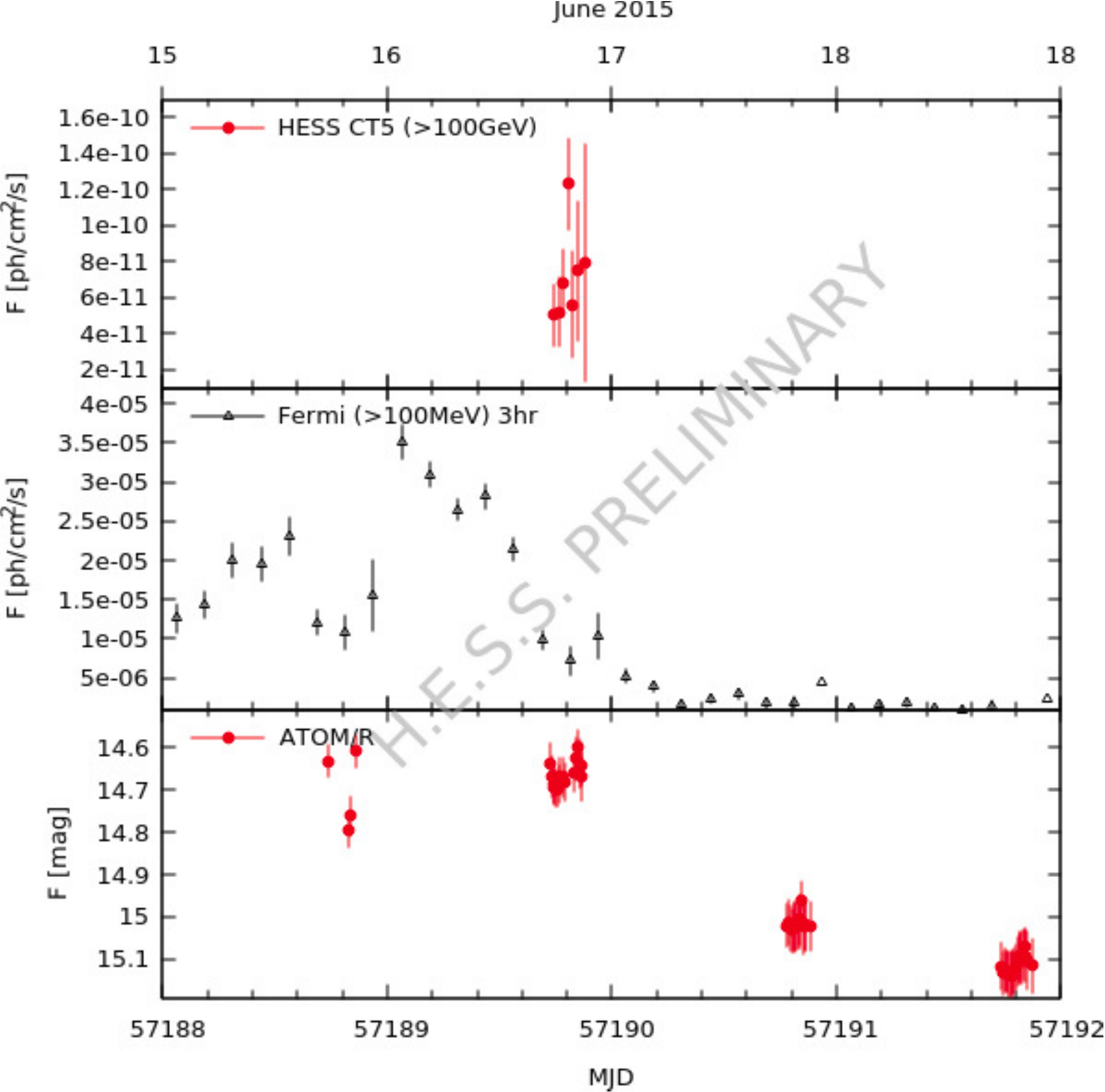}
\caption{Light-curves of the June 2015 flare of  \threec. From top to bottom: \hess\ integral flux above 100 GeV; \fermilat\ integral flux above 100 MeV and ATOM R magnitude. For reference, June 16, 2015 is MJD 57189. \label{3C279lc}}
\end{figure}

\begin{figure}
\includegraphics[width=250pt]{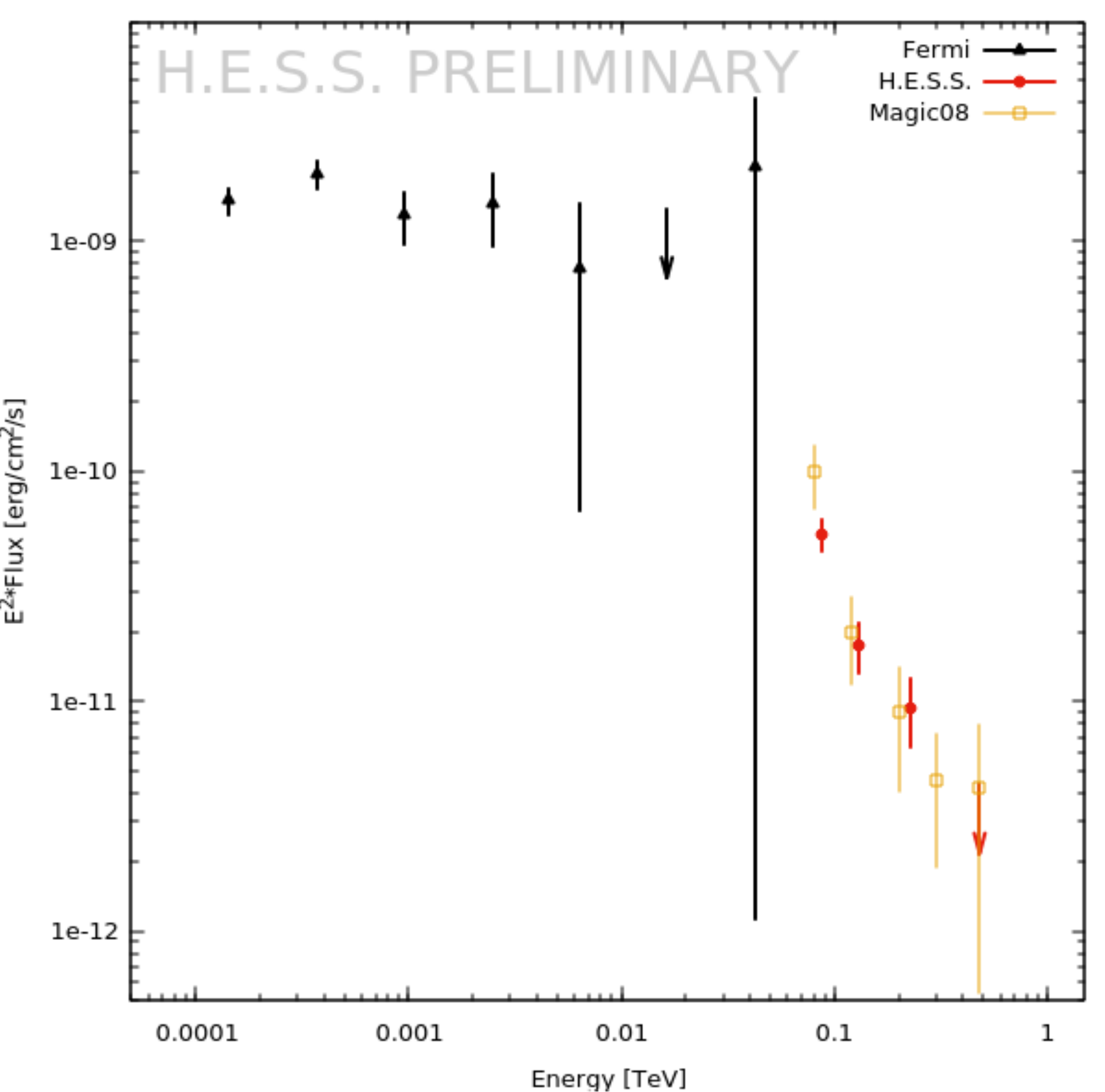}
\caption{Gamma-ray spectral energy distribution of \threec\ at the time of the \hess\ detection on June 16, 2015. Together with the preliminary \hess\ spectral point we present the \magic\ spectrum from \citep{MAGIC3C279}. The \fermilat\ spectrum is integrated over nine hours centred on the \hess\ observations. \label{3C279spec}}
\end{figure}

\threec\ is one of the brightest AGN in the $\gamma$-ray sky, and the first FSRQ ever detected at VHE, with the \magic\ telescopes, during the 2006 and 2007 outbursts \citep{MAGIC3C279, MAGIC3C279bis}. Afterwards, it was not detected again with Cherenkov telescopes, and during the 2011 $\gamma$-ray flare observed with \fermilat\ only VHE upper limits were presented by \magic\ and \veritas\ \citep{MAGIC3C279UL, VERITASUL}.\\

During June 2015, \threec\ entered a bright flaring state, as observed in the MeV-GeV energy band with \fermilat. Notably, this flare was the brightest ever observed from this source since the launch of the Fermi mission, and the first one for which minute timescale variability in the GeV regime is measured in a blazar \citep{3C279Fermi}. The \fermilat\ lightcurve, resulting from a re-analysis of \fermilat\ data, is shown in the middle plot of Fig. \ref{3C279lc}. The flux is binned in three hours intervals, and the highest state in the early hours of June 16 is evident.\\

The detection of the \fermilat\ flare triggered \hess\ ToO observations beginning on June 15. The \hess\ observing window lasted for about three hours each evening, and \threec\ was observed for five consecutive nights. Unfortunately, CT5 could only take part in the observations in the nights of June 16 and 18, while in the other nights only the small telescopes with a much higher energy threshold took observations. The analysis presented here focuses on the data taken with CT5 during the night of June 16, since these observations have the biggest chance of a successful detection due to the lower energy threshold, and since the $\gamma$-ray flare, as measured with \fermilat, was still ongoing.\\

The analysis of the \hess\ data results in the detection of VHE photons at a significance level of $8.7\ \sigma$ above the background, in 2.2 hours of live-time observations. The threshold of the analysis is estimated at $66$ GeV, at an average zenith angle of 27$^\circ$. The \hess\ lightcurve is provided in the top panel of Fig. \ref{3C279lc}. There is no statistically significant variability in the run-by-run (i.e. 28-minutes bins) light-curve. The spectrum is well described by a power-law with photon index $\Gamma=4.2 \pm 0.3$ and differential flux at 0.1 TeV equal to $(2.5 \pm 0.2)\times 10^{-9}$ cm$^{-2}$ s$^{-1}$ TeV$^{-1}$ (errors are statistical only). The preliminary \hess\ spectrum is shown in Fig. \ref{3C279spec}, together with the MeV-GeV spectrum by \fermilat\ (integrated over an interval of nine hours centred on the \hess\ observations on June 16). The \hess\ and \fermilat\ spectra are compatible with each other, and their respective extrapolations join smoothly. As a comparison, we also show the VHE spectrum of \threec\ as measured with the \magic\ telescopes in 2006 \citep{MAGIC3C279}. Interestingly, the 2006 and 2015 VHE spectra of \threec\ match each other, indicating that the source has been in a comparable state at these two epochs.\\

The bottom panel of Fig. \ref{3C279lc} shows the optical lightcurve of \threec\ as measured with the ATOM telescope. The optical trend seems to follow the $\gamma$-ray lightcurve measured with \fermilat. Similarly to the \hess\ observations, the peak of the \fermilat\ lightcurve during the morning of June 16, 2015, could not have been observed with \atom.\\

\section{\pks\ : a new quasar in the VHE sky}

\pks\ is an FSRQ, well studied at longer wavelengths since its discovery as a powerful radio source \citep{Day66}. Its optical/UV spectrum is characterized, as typical for FSRQs, by broad emission lines and the big-blue bump associated with the thermal emission from the accretion disk \citep{Malkan86}. Concerning its distance, \citep{Ho09} find a redshift $z=0.18941$. During January 2002, the source exhibited an extreme optical flare, with a variability of 0.6 magnitudes per hour \citep{Clements03}, which classifies the source as Optically Violently Variable (OVV) quasar. The source is detected in $\gamma$-rays with \fermilat, but since the launch of the \fermi\ mission it remained relatively calm. In November 2014, a first $\gamma$-ray flare was detected \citep{FermiAtel}. The activity continued in the following months, and a new bright $\gamma$-ray flare was seen in February 2015, which triggered the ToO \hess\ observations presented in this contribution.\\

\begin{figure}
\includegraphics[width=400pt]{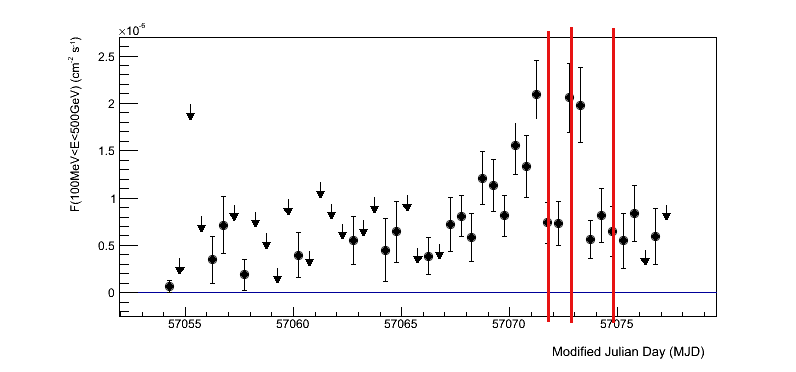}
\caption{\fermilat\ light-curve above 100 MeV of \pks. Every bin represents an interval of 12-hours. The time of \hess\ observations are marked in red. \label{0736fermi}}
\end{figure}

\begin{figure}
\includegraphics[width=500pt]{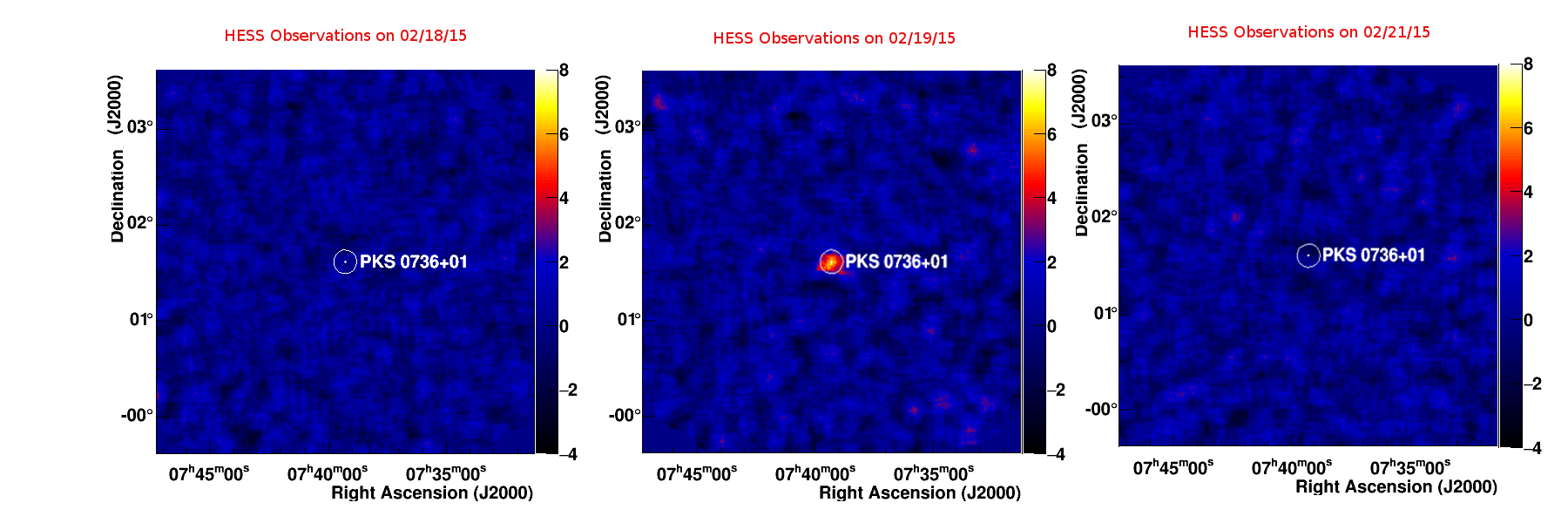}
\caption{\hess\ significance map centered on the position of \pks. From Left to Right: February 18,19, and 21, 2015. \label{0736hess}}
\end{figure}

The evolution of the flare as seen from \fermilat\ (integral flux from 100 MeV to 500 GeV, binned every 12 hours) is shown in Figure \ref{0736fermi}. The monitoring of the \fermilat\ observations detected an increased flux for the night of February 17, 2015, which triggered \hess\ ToO observations on February 18, 19 and 21. In Figure \ref{0736hess} we present the significance sky-maps, centered at the position of \pks, as reconstructed with the \hess\ monoscopic analysis described above. During the first night of \hess\ observations no significant VHE emission is detected from the source. On the other hand, on February 19, 2015, a clear signal at a signficance level of 7 $\sigma$ above the background is detected. The signal disappeared again two days later, on February 21. The preliminary flux estimation during the night of February 19, 2015, is about 10\% of the Crab Nebula flux above 100 GeV.\\

\pks\ is thus a new FSRQ in the VHE sky, the sixth detected so far and only the second discovered with \hess, after PKS 1510-089.\\ 

\newpage
\section{ACKNOWLEDGMENTS}
The support of the Namibian authorities and of the University of Namibia in facilitating the construction and operation of H.E.S.S. is gratefully acknowledged, as is the support by the German Ministry for Education and Research (BMBF), the Max Planck Society, the German Research Foundation (DFG), the French Ministry for Research, the CNRS-IN2P3 and the Astroparticle Interdisciplinary Programme of the CNRS, the U.K. Science and Technology Facilities Council (STFC), the IPNP of the Charles University, the Czech Science Foundation, the Polish Ministry of Science and Higher Education, the South African Department of Science and Technology and National Research Foundation, the University of Namibia, the Innsbruck University, the Austrian Science Fund (FWF), and the Austrian Federal Ministry for Science, Research and Economy, and by the University of Adelaide and the Australian Research Council. We appreciate the excellent work of the technical support staff in Berlin, Durham, Hamburg, Heidelberg, Palaiseau, Paris, Saclay, and in Namibia in the construction and operation of the equipment. This work benefited from services provided by the H.E.S.S. Virtual Organisation, supported by the national resource providers of the EGI Federation. N.C.  acknowledges  
support  from  Alexander  von  Humboldt  foundation.


\nocite{*}
\bibliographystyle{aipnum-cp}%
\bibliography{HESSTOO}%

\end{document}